\documentstyle[12pt]{article}
\input epsf
\begin{document}
\rightline{\vbox{\baselineskip12pt\hbox{EFI-95-70,
USC-95/028}\hbox{hep-th/9511052}}}
\vskip 1cm
\centerline{{\bf \Large \bf
Integrability in N=2 Gauge Theory: A Proof }}

\vskip 1cm
\centerline{{\bf \Large Emil J. Martinec}}
\vskip .4cm
\centerline{\it Enrico Fermi Institute and Department of Physics}
\centerline{\it University of Chicago,
5640 S. Ellis Ave., Chicago, IL 60637--1433 USA}

\vskip .4cm
\centerline{\it and}
\vskip .4cm
\centerline{{\bf \Large Nicholas P. Warner}}
\vskip .4cm
\centerline{\it Physics Department, University of Southern
California}
\centerline{\it University Park,  Los Angeles, CA 90089--0484 USA}


\def\pref#1{(\ref{#1})}

\def\ie{{i.e.}}
\def\eg{{e.g.}}
\def\cf{{c.f.}}
\def\etal{{et.al.}}
\def\etc{{etc.}}

\def\inbar{\,\vrule height1.5ex width.4pt depth0pt}
\def\IC{\relax\hbox{$\inbar\kern-.3em{\rm C}$}}
\def\IR{\relax{\rm I\kern-.18em R}}
\def\IP{\relax{\rm I\kern-.18em P}}
\def\Z{{\bf Z}}
\def\A{{\bf A}}
\def\B{{\bf B}}
\def\Pone{{\bf P^1}}
\def\gg{{\bf g}}
\def\hh{{\bf h}}
\def\ee{{\bf e}}
\def\aa{{\bf v}}
\def\h{{h_\gg}}
\def\hdual{{h_\gg^\vee}}
\def\One{{1\hskip -3pt {\rm l}}}
\def\nth{$n^{\rm th}$}
\def\adot{{\dot\alpha}}
\def\bdot{{\dot\beta}}
\def\R{{\RR}}
\def\ttwo{{{\bf T}^2}}
\def\zbar{{\bar z}}

\def\beq{\begin{equation}}
\def\eeq{\end{equation}}

\def\sst{\scriptscriptstyle}
\def\tst#1{{\textstyle #1}}
\def\frac#1#2{{#1\over#2}}
\def\coeff#1#2{{\textstyle{#1\over #2}}}
\def\half{\frac12}
\def\hf{{\textstyle\half}}
\def\ket#1{|#1\rangle}
\def\bra#1{\langle#1|}
\def\vev#1{\langle#1\rangle}
\def\d{\partial}

\def\np{{\it Nucl. Phys. }}
\def\pl{{\it Phys. Lett. }}
\def\pr{{\it Phys. Rev. }}
\def\ap{{\it Ann. Phys., NY }}
\def\prl{{\it Phys. Rev. Lett. }}
\def\mpl{{\it Mod. Phys. Lett. }}
\def\cmp{{\it Comm. Math. Phys. }}
\def\grg{{\it Gen. Rel. and Grav. }}
\def\cqg{{\it Class. Quant. Grav. }}
\def\ijmp{{\it Int. J. Mod. Phys. }}
\def\jmp{{\it J. Math. Phys. }}
\def\nextline{\hfil\break}
\catcode`\@=11
\def\slash#1{\mathord{\mathpalette\c@ncel{#1}}}
\overfullrule=0pt
\def\AA{{\cal A}}
\def\BB{{\cal B}}
\def\CC{{\cal C}}
\def\DD{{\cal D}}
\def\EE{{\cal E}}
\def\FF{{\cal F}}
\def\GG{{\cal G}}
\def\HH{{\cal H}}
\def\II{{\cal I}}
\def\JJ{{\cal J}}
\def\KK{{\cal K}}
\def\LL{{\cal L}}
\def\MM{{\cal M}}
\def\NN{{\cal N}}
\def\OO{{\cal O}}
\def\PP{{\cal P}}
\def\QQ{{\cal Q}}
\def\RR{{\cal R}}
\def\SS{{\cal S}}
\def\TT{{\cal T}}
\def\UU{{\cal U}}
\def\VV{{\cal V}}
\def\WW{{\cal W}}
\def\XX{{\cal X}}
\def\YY{{\cal Y}}
\def\ZZ{{\cal Z}}
\def\lam{\lambda}
\def\eps{\epsilon}
\def\vareps{\varepsilon}
\def\underrel#1\over#2{\mathrel{\mathop{\kern\z@#1}\limits_{#2}}}
\def\lapprox{{\underrel{\scriptstyle<}\over\sim}}
\def\lessapprox{{\buildrel{<}\over{\scriptstyle\sim}}}
\catcode`\@=12

\begin{abstract}
The holomorphic prepotential of ultraviolet finite
N=2 supersymmetric gauge theories is obtained by a
partial twisting of N=1 gauge theory in six dimensions,
compactified on $\IR^4\times\ttwo$.  We show that
Ward identities for the conserved chiral $\R$-symmetry
in these theories generate a set of constraints on the
correlation functions of chiral ring operators.
These correlators depend only on the coordinates of the
$\ttwo$, and the constraints are analogs of
the Knihnik-Zamolodchikov-Bernard equations
at the critical level.
\end{abstract}

\vskip .5cm

\section{Introduction}

There is by now a great deal of
evidence \cite{russians,MW,NT,DW,ejm,GM}
that integrability underlies the structure \cite{SW1-2}
of N=2 supersymmetric gauge theory.
In a recent article \cite{ejm}, the
first author made a number of remarks and conjectures
about the nature of this integrability.  Among these were
the following:\footnote{A discussion of some of these matters
from a rather different perspective may be found in \cite{GM}.}
\vskip .4cm

\noindent 1.)
An integrable system related to the
finite N=2 model with an adjoint hypermultiplet
(softly broken N=4 gauge theory)
was found by Donagi and Witten (explicitly for SU(N)),
in the context of Hitchin's integrable system \cite{hitchin}.
A more explicit but equivalent description of
the integrable model organizing
the effective theory is the elliptic Calogero-Moser model \cite{OP}
\beq
H_2=\hf p^2 + {\hf}\mu^2 \sum_{\alpha} \wp(\alpha\cdot q|\tau)\ ,
\label{CM-ham}
\eeq
where $\mu$ is the adjoint hypermultiplet mass, $\alpha$ are
the roots of the Lie algebra $\bf g$ of the gauge group $G$, and
$\wp$ is the Weierstrass function.
The microscopic coupling constant of the gauge theory appears in
the modulus of the torus
$\tau=\frac{\theta}{2\pi}+\frac{4\pi i}{e^2}$.
In the limit $\tau\rightarrow i\infty$,
$\mu\rightarrow\infty$, together with a shift of $\vec q$,
this integrable system degenerates to the
affine Toda lattice, which was found
in \cite{russians,MW,NT} to govern the pure N=2 gauge theory
that arises in the infrared limit.
\vskip .4cm

\noindent 2.)
Donagi and Witten \cite{DW} pointed out that the microscopic gauge
coupling $\tau$ should be part of the data that specifies the
integrable system, namely the modulus of the torus on
which the spectral parameter lives.
In ultraviolet finite N=2 theories, the
non-renormalization of the action allows precise enough control
over the theory that a proof of integrability should be possible.
A number of ideas about how this might transpire were
presented in \cite{ejm}.  In particular, it was proposed that
the requirement of finiteness should arise via anomaly
cancellation for the effective theory on the spectral parameter
torus, and that the periodicity of the potential \pref{CM-ham}
could arise from considerations of a higher-dimensional theory.
\vskip .4cm

\noindent 3.)
The partition function of finite N=2 gauge theories
should satisfy the Knihnik-Zamolodchikov-Bernard equation
on the spectral parameter torus of the integrable system,
at the critical level $k=-\hdual$.
The evidence for this is somewhat indirect.  The Toda system
that arises in the infrared limit of the softly broken
N=4 model is the {\it twisted} affine Toda lattice; its dynamics
comes from geodesic motion on the loop group $(LG)^\vee$,
projected down to a homogenous space by
gauging away a certain subgroup \cite{OP,AvM,GW}.
The appearance of the dual group in the infrared limit of
the finite theory accords with the idea \cite{MO} that the
effective theory should be a gauge theory of monopoles
with gauge group $G^\vee$.
In \cite{russians,MW,NT}, the partition function of the N=2 theory
was related to the WKB or Whitham-averaged Toda lattice,
which must come from the semiclassical $k\rightarrow\infty$ limit
of the loop group dynamics upstairs.
The appearance of the dual group in the infrared is fortuitous,
but in the ultraviolet one should see the microscopic gauge theory
with gauge group $G$; hence the Calogero-Moser model \pref{CM-ham}
should involve this group.  Remarkably, this interconversion
of UV group $G$ and IR group $G^\vee$ appears in work
on the quantization of the Hitchin integrable
system \cite{BD,FeFr}
(which, in turn, is intimately connected to the
model \pref{CM-ham} \cite{nekrasov,olsh}).
Namely, the semiclassical limit
$k^\vee\rightarrow\infty$ of the W-algebra $W_{k^\vee}(\gg^\vee)$
is the Gelfand-Dikii algebra $GD(\gg^\vee)$; this is nothing other
than the algebra of densities for the
conserved integrals of motion of the IR
integrable system.  This algebra is dual (in a well-defined sense)
to the critical level limit $k\rightarrow -\hdual$ of the
W-algebra $W_k(\gg)$, which is the algebra of commuting Hamiltonians
of the {\it quantized} Hitchin system -- whose
quadratic Hamiltonian on the torus
is the quantized version of \pref{CM-ham}.
Thus, Montonen-Olive duality would have its proper role in
supersymmetric gauge theory if the ultraviolet theory
were related to the {\it quantized} Calogero-Moser system
(or equivalently the quantized Hitchin system).
Interestingly, it would then also be intimately connected to
Langlands duality \cite{BD}, which is the context in which the
foregoing duality of W-algebras first arose.
\vskip .4cm

Our purpose here is to provide a proof of integrability in
finite N=2 gauge theories.  In the process, we will establish
much of the structure put forward in \cite{ejm}.
We will also give considerable substance to the tantalizing
parallels, described briefly in \cite{MW},
between the chiral ring of N=2 supersymmetric
models in two dimensions and the structure of the
correlators of chiral fields in N=2 supersymmetric QCD in
four dimensions.

The basic idea is remarkably simple.  Starting from N=1
gauge theory in six dimensions, we reduce to four dimensions
on a two-torus $\ttwo$ of modulus $\tau$ whose volume
is sent to zero.
Then, following \cite{BJSV},
we topologically twist the dynamics on the $\ttwo$ to yield
an effective dynamics in four-dimensional spacetime.
The four-dimensional theory has N=2 supersymmetry.
The BRST operators of the topological dynamics on $\ttwo$
are the analytic half of the N=2 supercharges in four dimensions
(the currents corresponding to the conjugate supercharges
become the BRST partner of the stress tensor on the $\ttwo$,
rendering the dynamics there trivial).
Thus the twisted theory computes the {\it holomorphic}
prepotential of the gauge theory -- any dependence on
antiholomorphic fields is BRST-trivial (up to
possible holomorphic anomalies \cite{BCOV}).
A holomorphic dependence on the coordinate $z$
of the $\ttwo$ remains.

A crucial feature of this construction is the fact that the
four-dimensional $\R$-symmetry comes from the local Lorentz symmetry
on the torus $\ttwo$.  The anomaly of the $\R$-symmetry may thus
be thought of as descending from the mixed Lorentz and gauge anomaly
in 6 dimensions.  Since the twisted theory is holomorphic on the
torus, the chiral Ward identity of the $\R$-symmetry is, from the
six-dimensional point of view, generated by the action of the
holomorphic energy-momentum tensor of the (conformal) field theory
on the torus $\ttwo$.
For suitable operators inserted on the $\ttwo$, this
action is in turn captured by the Knizhnik-Zamolodchikov-Bernard
(KZB) equations on the torus.  Thus, by using the chiral Ward identity
of the four-dimensional theory -- lifted to the six-dimensional
setting -- we discover the spectral curve,
and link the chiral correlators of the four-dimensional theory to
solutions of the (integrable) KZB equations.
While we have not fully evaluated the fermion correlation
functions which appear in these equations, we feel that
the method deserves a separate brief outline.  We will
complete the determination of these correlators, and
apply the resulting identities to various N=2 theories,
in a future work \cite{surely-you-jest}.

\section{Reduction from six dimensions}

The field content of N=1 gauge theory
in six dimensions \cite{BSS,sohnius}\footnote{We adopt
the gamma matrix conventions of Brink \etal\ \cite{BSS}.}
consists of the vector multiplet $(A_M,\lam_A)$ of gauge
fields and gauginos; together with a collection of matter
hypermultiplets $(\phi^i,\psi^i_{\bar A})$ and their conjugate
fields, transforming in some representations $R_i$ of the
gauge group $G$.  We denote six-dimensional
vector and spinor indices by $M$, $A$.
Under reduction on $\IR^4\times \ttwo$, we
denote the corresponding quantities $\mu$, $\alpha$, $\adot$
for $\IR^4$; and $m$, $a=\pm$ for $\ttwo$.
Irreducible spinors in six dimensions are Weyl spinors;
compatibility with a chiral supersymmetry requires that
$\lam_A$ be chiral, while $\psi_{\bar A}$ is antichiral.
We take the metric on the $\ttwo$ to be
\beq
ds^2=\frac{L^2}{\tau_2}|dz|^2\ ,
\eeq
where $dz=dx^4+\tau dx^5$.
In string theory $\tau=\tau_1+i\tau_2$ is the expectation value
of a gravitational vector multiplet $U$.  In addition we
will introduce a background antisymmetric tensor field
$B_{MN}$ with $\vev{B}=\frac{\theta}{2\pi}\eps_{56}$,
which may be considered as the real part of the vev of another
string-motivated vector multiplet $T$.  Then a
$B\wedge F\wedge F$ term in the action will induce the usual
theta term upon reduction to four dimensions.
For convenience we will take $T=U=\tau$ although this is
probably not essential.  One may then rather cavalierly
disregard, say, the dependence upon the theta angle
during the course of a derivation, knowing that it
will in the end be restored by holomorphicity.

The reduction on $\IR^4\times \ttwo$
splits the vector multiplet according to
\beq
(A_M,\lam_A)\rightarrow
(A_\mu,A_{++},A_{--};\lam_{\alpha+},\lam_{\adot-})\ .
\label{VM}
\eeq
Similarly, the hypermultiplet splits as
\beq
(\phi,\psi_{\bar A})\rightarrow
(\phi;\psi_{\adot+},\psi_{\alpha-})\ .
\label{HM}
\eeq
The four-dimensional theory has an $\R$-charge
which is nothing but the Lorentz spin on $\ttwo$
(later we will consider the two-dimensional
gauge fields $A_m$ to have
coordinate indices which do not transform under $\R$, and hence
are unaffected by topological twisting).

We now write the bosonic part of the field theory action:
\begin{eqnarray}
S&=&\frac{4\pi}{e^2}\int d^4x\; d^2z\;\left[
	\frac{L^2}{\tau_2}(F_{\mu\nu}F^{\mu\nu})+\frac{\tau_2}{L^2}
	(F_{mn}F^{mn})+(D_m A_\mu-D_\mu A_m)^2\right] +
\nonumber\\
& &\hskip 1cm
\frac{i\theta}{2\pi}\int d^4x\; d^2z\;
	\eps^{\mu\nu\rho\sigma}F_{\mu\nu}F_{\rho\sigma} +
\nonumber\\
& &
	\frac{4\pi}{e^2}\int d^4x\; d^2z\;\left[
	\frac{L^2}{\tau_2}(D_\mu \tilde\phi^i D^\mu \phi^i) +
	\frac{\tau_2}{L^2}(D_m \tilde\phi^i D^m\phi^i)
	+\frac{L^2}{\tau_2} M^j_{\ i}\tilde\phi^i\phi_j\right]\ .
\label{bos-action}
\end{eqnarray}
Now let us take the limit $L\rightarrow 0$,
$e\rightarrow 0$, $e/L$ fixed.  We implicitly rescale
$A_m$ by a factor of $L$ to keep it in the effective
theory.  This limit forces fields to be essentially flat
as far as their dependence on the coordinates $z$, $\zbar$
of the $\ttwo$.  The complex coupling constant of the
effective theory is $\tau$.
We may consider expanding
around a configuration with a Wilson line expectation
value
\beq
\vev{A_{--}}=v\cdot\hh\quad, \qquad
\vev{A_{++}}=\bar v\cdot\hh\ .
\eeq
In four-dimensional language, we have a non-zero Higgs vev;
on the $\ttwo$, we have a non-trivial flat gauge bundle.
For instance, holomorphic objects will satisfy
\beq
\OO(z+m+n\tau)=e^{-2\pi i (m+n\tau)
v\cdot\hh}\OO(z)e^{2\pi i (m+n\tau) v\cdot\hh}\ .
\eeq
As explained for instance in \cite{HMS}, all quantities
involving the Wilson line/Higgs
vev $v$ are invariant under the elliptic affine Weyl
transformations $\delta v \in \Lambda^\vee+\tau\Lambda^\vee$,
where $\Lambda^\vee$ is the coroot lattice.

Now consider twisting the theory.  Our discussion will
remain temporarily
at the level of the classical action; consideration
of the effects of quantum fluctuations are briefly
deferred.  A twist by the Lorentz spin along the $\ttwo$
has the effect
\begin{eqnarray}
\lam_{\alpha+}\rightarrow\lam_{\alpha}\quad,\qquad
\lam_{\adot-}\rightarrow\lam_{\adot--}
\nonumber\\
\psi_{\alpha-}\rightarrow\psi_{\alpha--}\quad,\qquad
\psi_{\adot+}\rightarrow\psi_{\adot}\ ;
\end{eqnarray}
similarly, the supersymmetry charges are shifted as
\begin{eqnarray}
Q_{\alpha+}\rightarrow Q_{\alpha}\quad,\qquad
Q_{\adot-}\rightarrow Q_{\adot--}
\nonumber\\
\bar Q_{\alpha+}\rightarrow \bar Q_{\alpha++}\quad,\qquad
\bar Q_{\adot-}\rightarrow \bar Q_{\adot}\ .
\end{eqnarray}
As is by now standard, we wish to reinterpret the
two-dimensional scalar
charge $Q_{\alpha}$ as a BRST operator\footnote{Alternatively,
one could choose $\bar Q_\adot$ as the BRST operator; however,
one cannot use both due to
$\{\bar Q_\adot,Q_\alpha\}=2P_{\adot \alpha}$.}; the current
associated to the conjugate charge $\bar Q_{\alpha++}$ will be
the BRST partner of the two-dimensional stress tensor.
Something remarkable has happened, though; the Weyl
condition in six dimensions, together with the topological
twist using the two-dimensional spin,
results in a set of BRST charges
which are analytic in four dimensions.
Having chosen the BRST operator to be $Q_{\alpha}$,
the effective action of the
twisted theory has as BRST invariant content
only the holomorphic part of the prepotential!
For instance, the supersymmetry transformation laws
\begin{eqnarray}
\delta A_{++}&=&\bar\zeta^\alpha_{++}\lam_\alpha  +
	\bar\lam^\alpha_{++}\zeta_\alpha
\nonumber\\
\delta A_{--}&=&\bar\zeta_\adot\lam^\adot_{--}  +
	\bar\lam_\adot\zeta^\adot_{--}
\end{eqnarray}
show that $A_{++}$ is BRST trivial, and so
the effective action of the twisted theory
only depends upon the holomorphic part $v$ of the
Higgs vev, and not on $\bar v$.

Apropos a proposal of \cite{ejm},
it is amusing to see that one can find a graded algebra
of charges in the twisted theory.  Namely, the contour
integrals $\oint A_{--}$, $\oint \psi^i_{\alpha--}$
are BRST invariant.  We do not
at present understand their utility.  However the even
charges are associated to the vector multiplets and the
odd ones to the hypermultiplets.

At this point we should discuss the effect of
quantum fluctuations on our theory.  We do not imagine
starting from the untwisted theory in six dimensions,
quantizing it, then twisting.  This would not make any
sense, as the original six-dimensional theory is sick in the
ultraviolet.  Rather, we wish to start with the twisted
theory on $\IR^4\times \ttwo$, and ask if it defines a
reasonable theory of four-dimensional fields, carrying
an additional holomorphic dependence on a parameter
$z$ which we may call the spectral parameter.
Fixing the gauge bundle over the $\ttwo$, the only
relevant field fluctuations are in $\IR^4$.
However, to derive the low-energy theory we made
a naive scaling analysis of the classical action.
This procedure is patently wrong in the quantum theory,
unless the gauge coupling $e^2$ does not undergo
anomalous scaling -- that is, we are dealing with
an ultraviolet finite N=2 model.

A separate argument leads to the same conclusion.
Consistently decoupling the non-holomorphic dependence
on the $\ttwo$ requires that the BRST charge
square to zero.  This implies a condition on the product
of two supersymmetry currents in the full six-dimensional
theory.  The conservation of supercurrents is related
by supersymmetry to the conservation of the stress tensor
on the $\ttwo$.
Both will be spoiled by a mixed anomaly\footnote{We thank
J. Harvey for a discussion on these matters.}
of the form
$tr\{R^2\}tr\{F^2\}$ in the six-dimensional theory,
where the $tr\{R^2\}$ comes from the $\ttwo$ and the $tr\{F^2\}$
from the $\IR^4$.  However $tr\{F^2\}$
will vanish in a theory whose matter
content corresponds to a finite N=2 theory in four dimensions.

\section{Ward identities}

In supersymmetric QCD in four dimensions, the correlators of
the lowest components of chiral superfields
satisfy supersymmetric Ward identities
which imply that these correlators are constant
({\it i.e.} independent of the location of the operators).  This
also remains true when instanton corrections are taken into account
(see, for example \cite{AKMRV}).  These operators
and their correlators may thus be thought of as defining
the chiral ring of the theory.
In terms of the six-dimensional theory, such
{\it topological} correlators will produce holomorphic conformal
blocks in the two-dimensional field theory on the torus.  As was
pointed out above, these will satisfy the KZB equation
on the once-punctured torus (\cf\ \cite{FW})
\beq
4\pi i(k+\hdual)\frac{\d}{\d\tau}\tilde\omega(z,\tau,\vec v)=
\biggl[\frac{\d^2}{\d{\vec v}^{\; 2}} -
\sum_\alpha \wp(\alpha\cdot v)\ee_{\alpha}\ee_{-\alpha} -
C_2(V)\eta_1(\tau) \biggr]\tilde\omega\ .
\label{KZB}
\eeq
In particular, as was conjectured in \cite{ejm}, the partition
function of the softly broken N=4 theory should satisfy this
equation in the critical level limit $k\rightarrow-\hdual$.
In this section we will start with the standard form
of the chiral symmetry Ward identity, and show how it can be
lifted into six dimensions as the KZB equation.
Consider the axial rotation
$\delta\psi(y)=-i\delta\eps(y)\gamma^5\psi(y)$ of a
single Dirac fermion $\psi$ of mass $M$
in representation $R$ of a background
gauge field $A_\mu$ (\cf\ \cite{coleman}):
\begin{eqnarray}
0&=&\d_\mu\vev{j^{\mu 5}(y)\OO_1(x_1)\ldots\OO_n(x_n)} +
\nonumber\\
& &\hskip 1cm
2M\vev{\bar\psi\gamma^5\psi(y)\OO_1(x_1)\ldots\OO_n(x_n)} +
\nonumber\\
& &\hskip 2cm
\sum_i\vev{\OO_1(x_1)\ldots\frac{\d\OO_i}{\d \eps}
\ldots\OO_n(x_n)} \delta(y-x_i) +
\nonumber\\
& &\hskip 3cm
\frac{iT_2(R)}{8\pi^2}
\vev{F\tilde F(y)\OO_1(x_1)\ldots\OO_n(x_n)}\ .
\end{eqnarray}
Integrating over $y$, and replacing the insertion of
the instanton number term by the corresponding derivative
with respect to the theta parameter, we find
\begin{eqnarray}
0&=& 2M\vev{\int d^4y \bar\psi\gamma^5\psi(y)
	\OO_1(x_1)\ldots\OO_n(x_n)} +
\nonumber\\
& &\hskip 1cm
\frac{\d}{\d\eps}\vev{\OO_1(x_1)\ldots\OO_n(x_n)} +
\nonumber\\
& &\hskip 2cm
4T_2(R)\frac{\d}{\d\theta}\vev{\OO_1(x_1)\ldots\OO_n(x_n)}\ .
\label{chiral-rot}
\end{eqnarray}
To apply this result to the twisted six-dimensional model above,
consider the relevant set of fermions \pref{VM},\pref{HM},
making opposite chiral rotations for vector
and hypermultiplet fields due to their opposite $\R$-charge.
Replace $\gamma^5$ in the first term by $\sigma^3\Gamma^7$.
The analog of the mass term $M$ comes from two sources:
The Higgs vev $v$ for all the fields, and the
flavor mass $M^i_{\ j}$ for the hypermultiplets
(which, if one wants, may be thought of as the vev of
a non-dynamical background flavor
gauge field).  Then \pref{chiral-rot} becomes
\begin{eqnarray}
\lefteqn{4\pi\bigl[2C_2(G)-\sum_i T_2(R_i)\bigr]\
\frac{\d}{\d\tau}\vev{\OO_1(x_1)\ldots\OO_n(x_n)} =}
\nonumber\\
& &\hskip 1cm
\vev{\int d^4y\biggl(\bar\lam^\alpha_{++} A_{--}\lam_\alpha +
\bar\psi^i_{\adot++} A_{--}\psi_i^{\adot}\biggr)
\OO_1(x_1)\ldots\OO_n(x_n)} +
\nonumber\\
& &\hskip 2cm
\vev{\int d^4y (M_{\sst --})^j_{\ i}
\bar\psi^i_{\adot++} \psi_j^{\adot}
\quad \OO_1(x_1)\ldots\OO_n(x_n)} +
\nonumber\\
& &\hskip 3cm
\sum_i\vev{\OO_1(x_1)\ldots\frac{\d\OO_i}{\d \eps}
\ldots\OO_n(x_n)}\ .
\label{ward}
\end{eqnarray}
We have anticipated that the result must be BRST invariant
by keeping only the holomorphic contributions on $\ttwo$.

Now let us interpret the various terms in \pref{ward}, and
compare with equation \pref{KZB}.
On the LHS we have the derivative with respect to the modulus
of the torus; encouragingly, its coefficient in \pref{ward}
vanishes for a finite theory -- the corresponding KZB-like
equation is at the `critical level'.  Of course, physically
this just means that the $\R$-symmetry is not violated
by instantons, and remains
unbroken in the full quantum theory.
Now specialize the operators $\OO_i$ to the chiral
ring of the four-dimensional N=2 theory.
These are fields whose correlators
have no $x$-dependence; however in general they will have
holomorphic $z$-dependence.
The last term in \pref{ward} then measures
the two-dimensional Lorentz spins of all the fields (recall that
the $\R$-symmetry transformation is just the Lorentz rotation
on the $\ttwo$), which for chiral ring operators is the same as
the holomorphic conformal dimension.  Thus two respective terms
in \pref{KZB} and \pref{ward} match,
since $C_2(V)$ in \pref{KZB} is just the holomorphic
conformal dimension of the operator at the puncture.

What about the other two terms?
After specializing to the chiral ring correlators,
all the terms in \pref{ward} are independent of $\IR^4$,
so we may interpret them as equations on the two-dimensional
spectral torus.  The first term on the
RHS may, using the equation of motion, be replaced by
$A_{--}\d_{++}^2 A_{--}$ (always up to BRST artifacts);
contour integrating over the location
$z$ of this operator on the $\ttwo$ gives $\oint (\d_{++}A_{--})^2$,
which is an insertion of the two-dimensional conjugate momentum
to $A_{--}$ squared.  A piece of this is $\d^2/\d v^2$.
Hence to identify the KZB equation, we have only to compute the
result of the flavor mass insertion -- the second term on the
RHS of \pref{ward}.  We confine ourselves here to a few
general comments.  The correlation function
\beq
\vev{\oint dz\int d^4y (M_{\sst --})^j_{\ i}
\bar\psi^i_{\adot++} \psi_j^{\adot}(y,z)
\ \OO_1(z_1)\ldots\OO_n(z_n)}
\label{mass}
\eeq
must be covariant under the shifts
$z\rightarrow z+m+n\tau$ and
$\vec v\rightarrow\vec v + \vec r_1 +\tau\vec r_2$
for $\vec r_1, \vec r_2\in \Lambda^\vee$.
It must have singularities when the $z_i$ collide.
These properties are shared by the KZB equation \pref{KZB}
and its multipuncture counterparts.
We expect them to arise from a careful evaluation of the
fermion propagator on the torus in the presence of
a background gauge field.
Of course, on general grounds, the Ward identity \pref{ward}
{\it must} be the Virasoro Ward identity on $\ttwo$, since
this is the effect of global chiral rotations on $\IR^4$
which are local Lorentz rotations on $\ttwo$.

Thus we have found a
partially topologically twisted N=1 gauge theory in
six dimensions which computes
the holomorphic prepotential of N=2 gauge theory
in four dimensions.  The dependence on the extra $\ttwo$
of the reduction is holomorphic, the extra variable becoming
a spectral parameter.
We also found a set of Ward identities satisfied by the
correlation functions of chiral ring operators in the theory
which bear a remarkable similarity to the
Knizhnik-Zamolodchikov-Bernard equations on $\ttwo$.
If these are not the KZB equations, they are what replace them
in the context of N=2 gauge theory.

\vskip 1cm
\noindent {\bf Acknowledgements:} We are grateful to
J. Harvey for discussions,
and to G. Moore and E. Witten for correspondence.
E.M. and N.W. are supported in part by funds
provided by the DOE under grant Nos.
DE-FG02-90ER-40560 and
DE-FG03-84ER-40168, respectively.

\end{document}